\documentclass{elsarticle}

\usepackage{amssymb,amsmath,stackrel,amsopn,latexsym}
\usepackage{graphicx}
\usepackage{amsfonts}
\usepackage{natbib}
\usepackage{color}

\usepackage{tikz}
\usetikzlibrary{arrows}
\usetikzlibrary{calc}
\newcommand{\dd}{1.2cm}
\usetikzlibrary{backgrounds}
\usetikzlibrary{fadings}
\usetikzlibrary{decorations.markings}
\usetikzlibrary{decorations.text}
\usetikzlibrary{decorations.shapes}
\usetikzlibrary{positioning}
\usetikzlibrary{fit}
\usetikzlibrary{shapes.arrows}

\definecolor{TUEblue}{rgb}{0,0.4,0.8}
\definecolor{darkgreen}{rgb}{0.13, 0.55, 0.13}

\begin{document}

\begin{frontmatter}
\title{Identification in Dynamic Networks\tnoteref{footnoteinfo}} 

\tnotetext[footnoteinfo]{Accepted for publication in Computer and Chemical Engineering, 2017. A preliminary version of this paper was presented at the Chemical Process Control Conference  (FOCAPO/CPC 2017), Tucson, AZ, January 8-12, 2017.}

\author[First]{Paul M.J. Van den Hof \corref{cor1}}
\author[Second]{Arne G. Dankers}
\author[First]{Harm H.M. Weerts}

\cortext[cor1]{Corresponding author}
\address[First]{Control Systems Group, Department of Electrical Engineering, Eindhoven University of Technology, The Netherlands (email: h.h.m.weerts@tue.nl, p.m.j.vandenhof@tue.nl)}
\address[Second]{Department of Electrical Engineering, University of Calgary, Canada (email: adankers@hifieng.com)}

\begin{keyword}
system identification, dynamic networks, identifiability, experiment design, model-based control, distributed control, closed-loop identification.
\end{keyword}

\begin{abstract}
System identification is a common tool for estimating (linear) plant models as a basis for model-based predictive control and optimization. The current challenges in process industry, however, ask for data-driven modelling techniques that go beyond the single unit/plant models. While optimization and control problems become more and more structured in the form of decentralized and/or distributed solutions, the related modelling problems will need to address structured and interconnected systems. An introduction will be given to the current state of the art and related developments in the identification of linear dynamic networks. Starting from classical prediction error methods for open-loop and closed-loop systems, several consequences for the handling of network  situations will be presented and new research questions will be highlighted.
\end{abstract}
\end{frontmatter}

\section{Introduction}
\noindent
System identification is a well-developed technology for estimating plant models from operational data, typically taken during dedicated plant testing/excitation. Data-driven estimation and maintenance of dynamic models is considered a key technology for realizing a higher level of autonomy of model-based controllers when maintaining economic optimal operation of the plant, see e.g. \cite{Gevers:05,Zhu:06,Darby&Nikolaou:14,Autoprofit:16}.

The system configurations that are typically being considered are either multivariable open-loop or feedback controlled (closed-loop) systems. Whereas in open-loop identification the plant input signals are not restricted by the system, in closed-loop systems the presence of feedback induces a correlation of the plant's output disturbances with the plant input, thereby complicating the identification problem, see e.g. \cite{ljung:99}. This has led to the development of dedicated closed-loop identification schemes  \citep{VandenHof&Schrama:95,Forssell&Ljung:99}

In several areas of technology, the development of controlling and optimizing system's operations, involves the handling of structure and of interacting subsystems. This is happening in solutions for decentralized and distributed process control, see e.g. \cite{Rawlings&Stewart:08,Christofides&etal:13}. Also in other technology domains, like power networks and robotic networks, interconnection structures are playing an increasing role, while in areas like systems biology the modelling and identification of interconnected systems, including the topology, is a key problem.

\bigskip
\begin{figure}[h]
\centering
\includegraphics[width=.65 \linewidth]{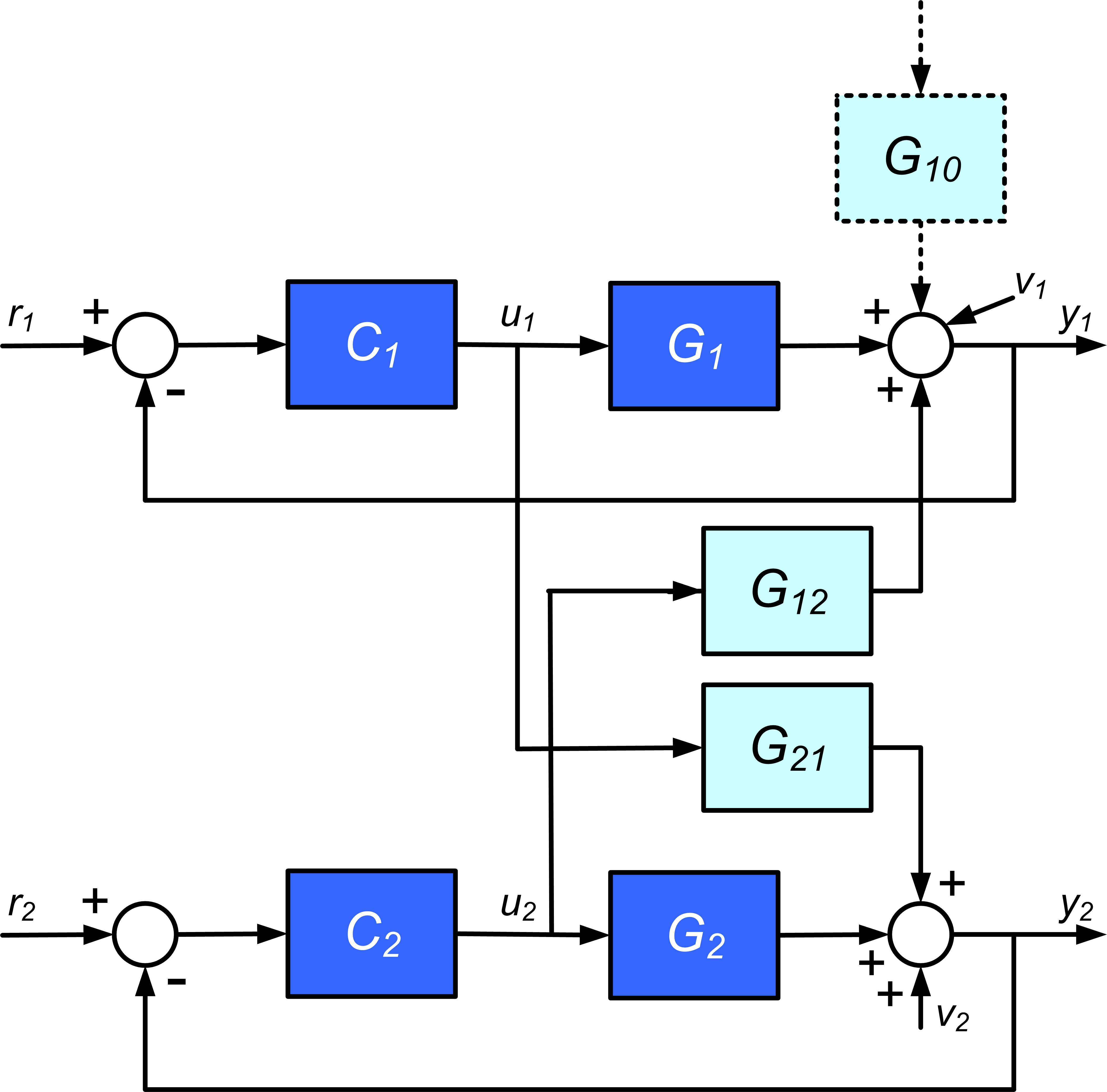}
\caption{Interacting dynamics in two control loops}\label{figure:3}
\end{figure}

A nice example of a structured identification problem is the problem of two interconnected controlled systems, as present in Figure \ref{figure:3}, which is considered in \cite{Gudi&Rawlings:06}, for the particular situation that $G_{12}=0$, and where the identification problem is to identify the interacting dynamics $G_{21}$ and possibly $G_{12}$. The handling of such structured systems should be facilitated by a theory for identification in dynamic networks, which is lacking in the classical identification literature. In this paper several steps in the recent development of such a theory are highlighted and illustrated.

\section{Linear dynamic networks}

Following the basic setup of \cite{VandenHof&etal:13}, a dynamic network is built up out of $L$ scalar \emph{internal variables} or \emph{nodes} $w_j$, $j
= 1, \ldots, L$, and $K$ \emph{external variables} $r_k$, $k=1,\cdots K$.
Each internal variable is described as:
\begin{align}
w_j(t) = \sum_{\stackrel{l=1}{l\neq j}}^L
G_{jl}^0(q)w_l(t) + \sum_{k=1}^L F^0_{jk}(q)r_k(t) + v_j(t)
\label{eq:netw_def}
\end{align}
where $q^{-1}$ is the delay operator, i.e. $q^{-1}w_j(t) = w_j(t-1)$;
\begin{itemize}
	\item $G_{jl}^0$, are proper rational transfer functions, and the single transfers $G_{jl}^0$ are referred to as {\it modules} in the network.
	\item $r_k$ are \emph{external variables} that can directly be manipulated by the user. Without loss of generality we will assume in this paper that $F^0_{jk} = 0$, for $j\neq k$, and $F^0_{jj} \in \{0,1\}$, implying that - when present - the external signal $r_j$ directly affects $w_j$.
	\item $v_j$ is \emph{process noise}, where the vector process $v=[v_1 \cdots v_L]^T$ is modelled as a stationary stochastic process with rational spectral density, such that there exists a $p$-dimensional white noise process $e:= [e_1 \cdots e_p]^T$ with diagonal
covariance matrix $\Lambda^0>0$ and $p \leq L$,  such that
$v(t) = H^0(q)e(t)$,
with $H^0$ a proper rational transfer function matrix that is monic and stable and has a stable left inverse.\\
\end{itemize}

\begin{figure}[h]
\centering
\includegraphics[width=.50 \linewidth]{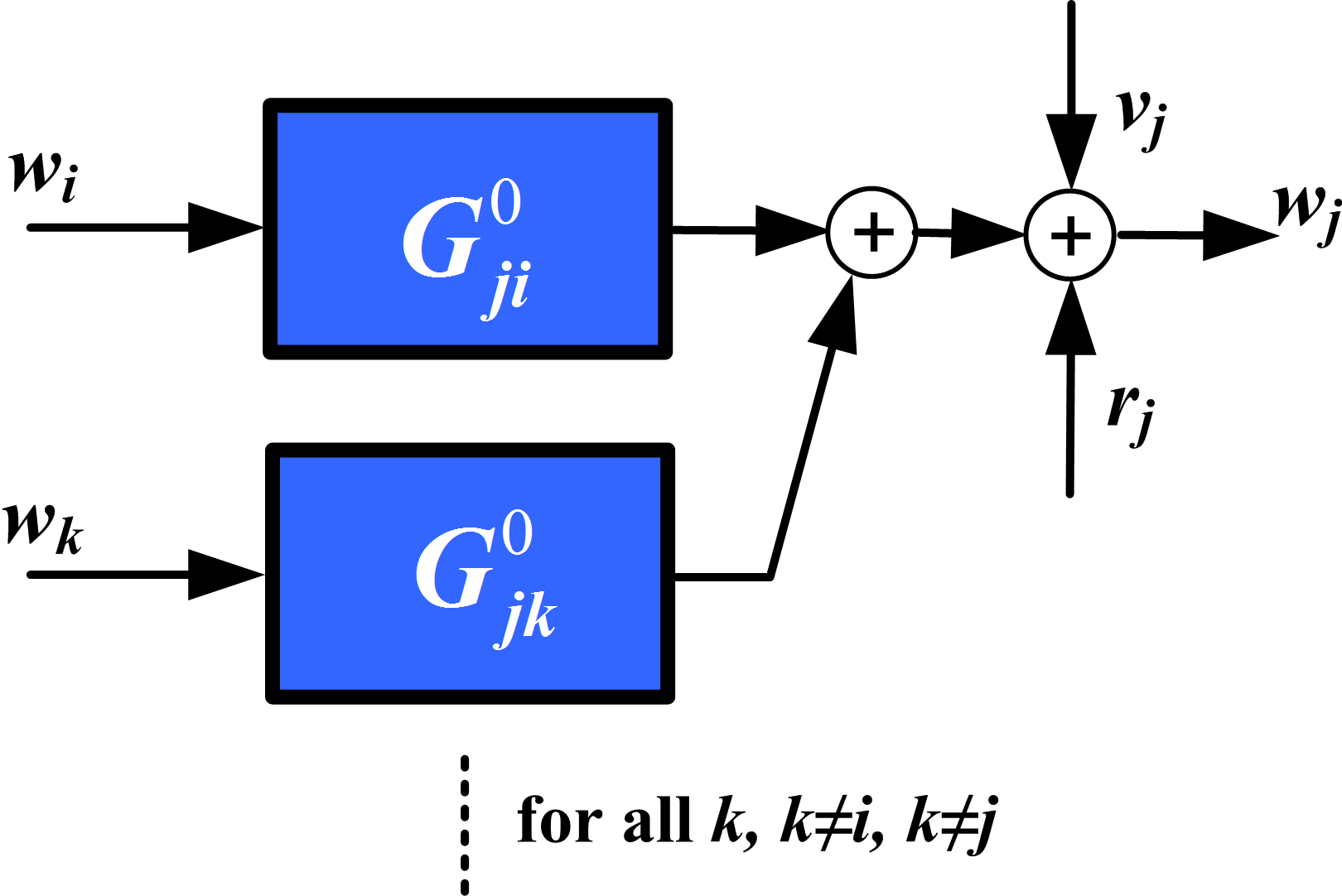}
\caption{Node building block of a network}\label{figure:2}
\end{figure}

A single building block of the network is depicted in Figure \ref{figure:2}.

The situation that we would like to consider is the full network constructed by combining (\ref{eq:netw_def}) for all node signals,
\begin{align*}
\begin{bmatrix}  \! w_1 \!  \\[7pt] \! w_2 \!  \\[1pt]  \! \vdots \! \\[1pt] \! w_L \!  \end{bmatrix} \!\!\! = \!\!\!
\begin{bmatrix}
0 &\! G_{12}^0 \!& \! \cdots \! &\!\! G_{1L}^0 \!\\
\! G_{21}^0 \!& 0 & \! \ddots \! &\!\!  \vdots \!\\
\vdots &\! \ddots \!& \! \ddots \! &\!\! G_{L-1\ L}^0 \!\\
\! G_{L1}^0 \!&\! \cdots \!& \!\! G_{L\ L-1}^0 \!\! &\!\! 0
\end{bmatrix} \!\!\!\!
\begin{bmatrix} \! w_1 \!\\[7pt]  \! w_2 \!\\[1pt] \! \vdots \!\\[1pt] \! w_L \! \end{bmatrix} \!\!\!
+ \!\! F^0 \!
\begin{bmatrix} \! r_1 \!\\[7pt] \! r_2 \!\\[1pt] \! \vdots \!\\[1pt]  \! r_{L} \!\end{bmatrix}
\!\!\!+\!\!
H^0 \! \begin{bmatrix}\! e_1 \!\\[7pt] \! e_2 \!\\[1pt] \! \vdots \!\\[1pt] \! e_p\!\end{bmatrix} \!\!\!
\end{align*}
Using obvious notation this results in the matrix equation:
\begin{align} \label{eq.dgsMatrix}
w = G^0(q)w + R^0(q)r + H^0(q)e,
\end{align}
where $R^0(q)$ is the $L \times K$ submatrix of $F^0$ composed of those columns of $F^0$ that relate to external variables that are actually present, while the present excitation signals are collected in the $K$-dimensional vector signal $r$.

\begin{figure}
\centering
\includegraphics[width=\linewidth]{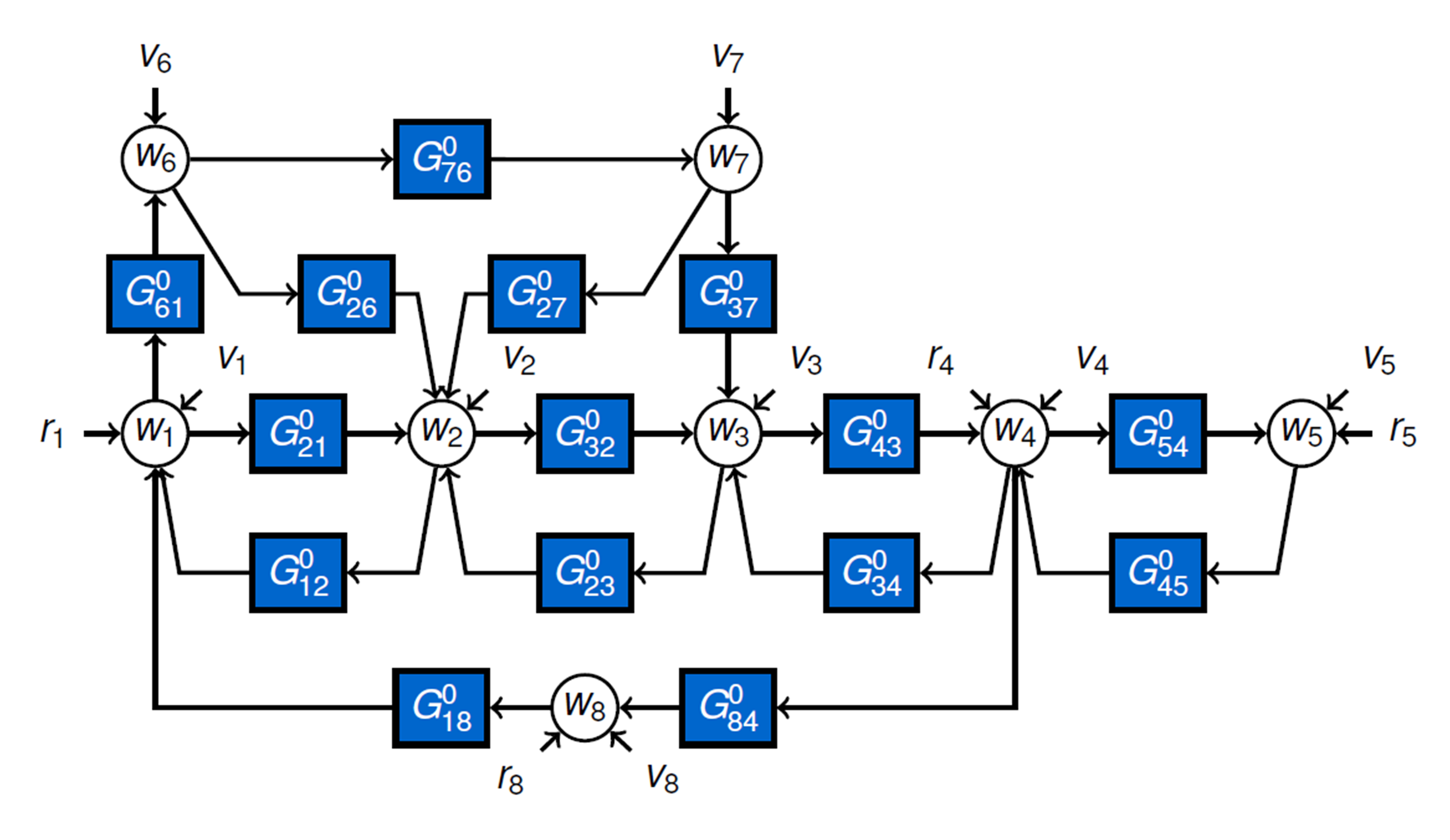}
\caption{Example of a dynamic network}\label{figure:1}
\end{figure}

\section{Identification of a single module - the full MISO approach}
\subsection{Direct method}
\label{secdir}
When the objective is to identify a single module in the network, denoted as $G^0_{ji}$, while the topology of the network is known, there is a direct identification algorithm that can provide a consistent estimate of this module dynamics. In order to apply this method the following additional assumptions are formulated:\\
$\bullet$\ The spectral density $\Phi_v(\omega)$ is diagonal, i.e. all noise signals are mutually uncorrelated;\\
$\bullet$\ Every loop around node signal $w_j$ has a delay (no algebraic loops);

Next we will formulate the algorithm, which is a direct generalization of the classical direct method for closed-loop identification (\cite{VandenHof&etal:13}):

\begin{enumerate}
\item Determine the set $\mathcal{N}_j$ of all node numbers $k$ such that $G^0_{jk}\neq 0$; all node signals in this set need to be measured, besides $w_j$;
\item Determine the subset $\mathcal{K}_j \subset \mathcal{N}_j$ of all node numbers $k$ such that $G^0_{jk}$ is known a priori;
\item Denote $\mathcal{U}_j = \mathcal{N}_j\backslash\mathcal{K}_j$, as the set of node signals $k$ for which $G^0_{jk}$ needs to be estimated;
\item Determine $\bar w_j(t) := w_j(t)-r_j(t)- \sum_{k\in \mathcal{K}_j} G_{jk} w_k(t)$
\item Solve the identification problem
\[ \hat\theta_N := \arg\min_{\theta} \frac{1}{N} \sum_{t=0}^{N-1} \varepsilon^2(t,\theta),\ \ \mbox{with} \]
\begin{equation} \varepsilon(t,\theta) := H_j(\theta)^{-1}[\bar w_j(t)-
   \sum_{k\in \mathcal{U}_j} G_{jk}(\theta) w_k(t)].\label{eq:pe1} \end{equation}
\end{enumerate}

Then $G_{jk}(\hat\theta_N)$, $k \in \mathcal{U}_j$ and $H_j(\hat\theta_N)$ are estimated consistently, provided that the model set is parametrized so as to contain the real underlying system (system in the model set), and the input signals of the estimated modules $G_{jk}(\theta)$ are sufficiently informative.\footnote{More detailed conditions for informativity of the data are explored in \cite{Gevers&Bazanella:15}.}

When applying this algorithm to the example network in Figure \ref{figure:1}, in a situation that the  objective would be to identify the module $G^0_{21}$, it follows that $\mathcal{N}_2 = \{1,3,6,7\}$. If all the corresponding modules are unknown, they need to be identified, leading to a 4-input, 1-output identification problem. The consistent estimation of the target module $G^0_{21}$ is then embedded in the consistent estimation of all $4$ modules in the considered MISO system.

\subsection{Two-stage / projection approach}
\label{sec2s}
Whereas the direct identification method requires exact noise modelling in order to arrive at consistent module estimates, a second approach, called the two-stage or projection approach, allows to estimate module dynamics independently from noise models. Rather than using measured node signals as model inputs, this method uses projected node signals, i.e. node signals that are projected onto external excitation signals $r_i$, as introduced in \cite{VandenHof&Schrama:93} for classical closed-loop identification problems.

The algorithm from the previous subsection is now adapted as follows:

\begin{enumerate}
\item Select a set of excitation signals $\{r_m\}$, with $m \in \mathcal{R}_{is}$ that are correlated with $w_i$.
\item Denote $\mathcal{U}_{is} \subset \mathcal{U}_j$ as the set of node signals that is correlated to any of the excitation signals in $\mathcal{R}_{is}$.
\item Determine the projected signals $w_k^{(\mathcal{R}_{is})}$, for $k \in \mathcal{U}_{is}$. This can be done through correlation techniques, or through an estimation procedure, according to \cite{VandenHof&Schrama:93}.
\item Proceed with steps 4-5 above, with the prediction error
\[ \varepsilon(t,\theta) := H_j(\eta)^{-1}[\bar w_j(t)-
   \sum_{k\in \mathcal{U}_{is}} G_{jk}(\theta) w_k^{(\mathcal{R}_{is})}(t)]. \]
\end{enumerate}

Then $G_{jk}(\hat\theta_N)$, $k \in \mathcal{U}_{is}$ are estimated consistently, provided that the model set is parametrized so as to contain the real underlying system (modules in the model set), and the projected input signals $w_k^{(\mathcal{R}_{is})}$ are sufficiently informative.\footnote{For the two-stage / projection approach the condition on absence of algebraic loops around $w_j$ can be removed.}

When applying this algorithm to the example of Figure \ref{figure:1}, and focussing again on estimating module $G^0_{21}$, then, when choosing the external signal $r_1$, so $\mathcal{R}_{is} =\{1\}$, it follows that $\mathcal{U}_{is} = \{ 1,3,6,7\}$. So again we will have a 4-input, 1-output identification problem. However when all input signals are projected onto $r_1$, the resulting signals might not be sufficiently informative. In order to reach this, an option to include the external signals $r_4, r_5$ and $r_8$ too, leading to $\mathcal{R}_{is} =\{1,4,5,8\}$. The network topology conditions on the excitation signals $r_m$, i.e. the possible correlations with node signals, can be verified by algorithms from graph theory \citep{VandenHof&etal:13}.

\section{Predictor input selection}
\subsection{Direct method}
The methods discussed in the previous section include basically all node signals in $\mathcal{N}_j$ as inputs in the  predictor models. Including all possible inputs is typically not necessary; there is an opportunity to make a more sparse selection. This can be of importance if some node signals in the network are hard (or expensive) to measure. The methods presented below originate from \cite{Dankers&etal_TAC:16}.

We are going to construct a set $\mathcal{D}_j \subset \mathcal{U}_j$, being a set of node variables that will serve as predictor inputs.\footnote{It is actually not strictly necessary that $\mathcal{D}_j \subset \mathcal{U}_j$.}

In view of the consistent identification of the module $G_{ji}^0$, a relaxed set of conditions that needs to be satisfied for this set $\mathcal{D}_j$ can be formulated as follows: \\
(a) $i \in \mathcal{D}_j$, $j \notin \mathcal{D}_j$;\\
(b) every path from $w_i$ to $w_j$, excluding the path $G^0_{ji}$, goes through a node $w_k$, $k\in \mathcal{D}_j$;\\
(c) every loop through $w_j$ goes through a node $w_k$, $k\in \mathcal{D}_j$.

These conditions state that every path that is parallel to $G^0_{ji}$ and every loop around $w_j$ should be ``blocked'' by a predictor input. If these conditions are satisfied for $\mathcal{D}_j$, then a reduced (immersed) network can be constructed composed of node signals $\{j,\mathcal{D}_j\}$. The immersed network is a network in which a particular set of nodes is removed, and therefore a network with a reduced set of node signals remains. However the node signals that remain are invariant, i.e. they are exactly the same signals as in the original network. This immersed network will have module dynamics $\breve G^0_{jk}$, $k \in \mathcal{D}_j$, that in general will be different from the original module dynamics, while the node signals remain invariant. However under the conditions listed above it holds that $G^0_{ji} = \breve G^0_{ji}$, and therefore in the immersed network we can still identify the intended module $G^0_{ji}$.

In order to guarantee consistency of the module estimate, one additional condition needs to be satisfied, which is related to the notion of confounding variables.

In the current setup of output variable $w_j$ and a set $\mathcal{D}_j$ of predictor inputs, a variable $v_\ell$ is called {\it a confounding variable} if it directly affects the output $w_j$ as well as at least one of the inputs $w_k$, $k \in \mathcal{D}_j$. Formally stated, $v_\ell$ is a confounding variable if there exists an input node $k \in \mathcal{D}_j$ such that there exist paths from $v_\ell$ to $w_k$ and from $v_\ell$ to $w_j$ that do not pass through a node in $\{j,\mathcal{D}_j\}$. Confounding variables are non-measured variables that affect both input and output, and create correlation between the two signals that is not generated by the dynamics of the module of interest.

\medskip
After construction of the set $\mathcal{D}_j$ the direct MISO identification method of section \ref{secdir} can be applied with the predictor input set $w_k, k\in\mathcal{D}_j$. Under the usual conditions, listed in Section \ref{secdir}, the module $G^0_{ji}$ is estimated consistently provided that no disturbance signal $v_\ell$ is a confounding variable.

\medskip
If we return to the example of Figure \ref{figure:1} and the modelling of $G^0_{21}$, it follows that in order to satisfy conditions (a)-(c) above, it would suffice to choose $\mathcal{D}_j = \{1, 3, 6\}$. In comparison with the previous section, node signal $w_7$ is not necessary as a predictor input for the immersed network to maintain $G^0_{21}$. However in this setting, $w_7$ now acts as a confounding variable, since it has a path to input $w_3$ and output $w_2$. This confounding variable can be ``blocked'' by including $w_7$ as a predictor input into $\mathcal{D}_j$, as in that situation $w_7$ is no longer a confounding variable. As a result, we have not been able to effectively reduce the set of predictor inputs in this particular case, in comparison with the situation described in Section \ref{secdir}.

\subsection{Two-stage / projection approach}

A similar result as in the previous section can be formulated for the two-stage / projection approach, but with slightly varying conditions. The most important difference being that for this method the presence of confounding variables is no problem for consistency. The selection of input nodes $\mathcal{D}_j$ needs to satisfy conditions (a)-(c) from the previous section, in order to guarantee the correct dynamics being present in $\breve G^0_{ji}$.

Let $\{r_m\}, m \in \mathcal{T}_j$ be the external excitation signals onto which the input node signals $\{w_k\},k \in \mathcal{D}_j$ will be projected, while this projection is denoted as  $w_k^{(\mathcal{T}_j)}$.

\medskip
Then application of the two-stage / projection method, on the basis of the prediction error:
\[ \varepsilon(t,\theta) := \breve H_j(\eta)^{-1}[\bar w_j(t)-
   \sum_{k\in \mathcal{D}_{j}} \breve G_{jk}(\theta) w_k^{(\mathcal{T}_{j})}(t)] \]
will lead to a consistent estimate of $\breve G^0_{ji} = G^0_{ji}$, under conditions that are similar to the conditions as formulated in Section \ref{sec2s}, while additionally there should not exist external signals that are selected in $\mathcal{T}_j$ that have paths to $w_j$ that do not pass through nodes in $\mathcal{D}_j$. Such external excitation signals would then act as a disturbance on the output that is correlated with the inputs.

\begin{figure}[bht]
\begin{center}
\scalebox{0.85}{
\begin{tikzpicture}
[tf/.style={rectangle,fill=TUEblue, draw, very thick, text=white, minimum height=0.4cm, minimum width=0.5cm, text centered},
 sum/.style={circle,draw,minimum size=0.25cm, thick, inner sep =1pt},
hbox/.style={rectangle,fill=fadedyellow, draw, very thick},
 node distance = 0.5cm, line width = 1pt]
\node[sum,fill=darkgreen] (w1){$w_1$};
\node[tf] (g21)[right=of w1]{$G_{21}^0$};
\node[tf,fill=green,text=black] [right=of w1]{$G_{21}^0$};
\node[sum,fill=darkgreen] (w2)[right=of g21]{$w_2$};
\node[tf] (g32)[right=of w2]{$G_{32}^0$};
\node[sum,fill=darkgreen] (w3)[right=of g32]{$w_3$};
\node[tf] (g43)[right=of w3]{$G_{43}^0$};
\node[sum] (w4)[right=of g43]{$w_4$};
\node[tf] (g54) [right=of w4] {$G_{54}^0$};
\node[sum] (w5) [right=of g54] {$w_5$};
\node[tf] (g45) [below=of g54] {$G_{45}^0$};
\node[tf] (g12)[below=of g21]{$G_{12}^0$};
\node[tf] (g23)[below=of g32]{$G_{23}^0$};
\node[tf] (g34)[below=of g43]{$G_{34}^0$};
\node[sum] (w8)[below=of g23]{$w_8$};
\node[tf] (g84)[right=of w8]{$G_{84}^0$};
\node[tf] (g18)[left=of w8]{$G_{18}^0$};
\node[tf] (g26)[above=of g21,xshift=0.4cm]{$G_{26}^0$};
\node[tf] (g61) at (g26 -| w1){$G_{61}^0$};
\node[tf] (g27)[above=of g32,xshift=-0.4cm]{$G_{27}^0$};
\node[tf] (g37) at (g27 -| w3){$G_{37}^0$};
\node[sum,fill=darkgreen] (w6)[above=of g61]{$w_6$};
\node[sum] (w7)[above=of g37]{$w_7$};
\node[tf] (g76) at ($0.5*(w6) + 0.5*(w7)$){$G_{76}^0$};

\node (v1)[above right=of w1,yshift=-0.2cm,xshift=-0.2cm]{$v_1$};
\node (v2)[above right=of w2,yshift=-0.2cm,xshift=-0.2cm]{$v_2$};
\node (v3)[above right=of w3,yshift=-0.2cm,xshift=-0.2cm]{$v_3$};
\node (v4)[above right=of w4,yshift=-0.2cm,xshift=-0.2cm]{$v_4$};
\node (v5)[above right=of w5,yshift=-0.2cm,xshift=-0.2cm]{$v_5$};
\node (v6)[above=of w6,yshift=-0.2cm]{$v_6$};
\node (v7)[above=of w7,yshift=-0.2cm]{$v_7$};
\node (v8)[below right=of w8,yshift=0.2cm,xshift=-0.2cm]{$v_8$};

\node (r1)[left=of w1,xshift=0.2cm]{$r_1$};
\node (r4)[above left=of w4,xshift=0.2cm,yshift=-0.2cm]{$r_4$};
\node (r5)[right=of w5,xshift=-0.2cm]{$r_5$};
\node (r8)[below left=of w8,xshift=0.2cm,yshift=0.2cm]{$r_8$};

\draw[->] (v1) -- (w1);
\draw[->] (v2) -- (w2);
\draw[->] (v3) -- (w3);
\draw[->] (v4) -- (w4);
\draw[->] (v5) -- (w5);
\draw[->] (v6) -- (w6);
\draw[->] (v7) -- (w7);
\draw[->] (v8) -- (w8);
\draw[->] (r1) -- (w1);
\draw[->] (r4) -- (w4);
\draw[->] (r5) -- (w5);
\draw[->] (r8) -- (w8);

\draw[->] (w1) -- (g21);
\draw[->] (g21) -- (w2);
\draw[->] (w2) -- (g32);
\draw[->] (g32) -- (w3);
\draw[->] (w3) -- (g43);
\draw[->] (g43) -- (w4);
\draw[->] (w4) -- (g54);
\draw[->] (g54) -- (w5);
\draw[->] (w4) |- (g84);
\draw[->] (g84) -- (w8);
\draw[->] (w8) -- (g18);
\draw[->] (g18) -| (w1);
\draw[->] (w1) -- (g61);
\draw[->] (g61) -- (w6);
\draw[->] (w6) -- (g76);
\draw[->] (g76) -- (w7);
\draw[->] (w7) -- (g37);
\draw[->] (g37) -- (w3);

\draw[->] (w4) -- ($(w4 |- g34)+(-0.2cm,0)$) -- (g34);
\draw[->] (g34) -- ($(g34 -| w3)+(0.2cm,0)$) -- (w3);
\draw[->] (w3) -- ($(w3 |- g23)+(-0.2cm,0)$) -- (g23);
\draw[->] (g23) -- ($(g23 -| w2)+(0.2cm,0)$) -- (w2);
\draw[->] (w2) -- ($(w2 |- g12)+(-0.2cm,0)$) -- (g12);
\draw[->] (g12) -- ($(g12 -| w1)+(0.2cm,0)$) -- (w1);
\draw[->] (w5) -- ($(w5 |- g45)+(-0.2cm,0)$) -- (g45);
\draw[->] (g45) -- ($(g45 -| w4)+(0.2cm,0)$) -- (w4);

\draw[->] (w6) -- ($(w6 |- g26)+(0.7cm,0)$) -- (g26);
\draw[->] (g26) -- ($(g26 -| w2)+(-0.2cm,0)$) -- (w2);
\draw[->] (w7) -- ($(w7 |- g27)+(-0.7cm,0)$) -- (g27);
\draw[->] (g27) -- ($(g27 -| w2)+(0.2cm,0)$) -- (w2);
\end{tikzpicture}}

\end{center}
\caption{Network with selected predictor inputs for two-stage identification of $G^0_{21}$. The selected inputs are projected onto  $r_1, r_4, r_5, r_8$.\label{fig.2sinput}}
\end{figure}
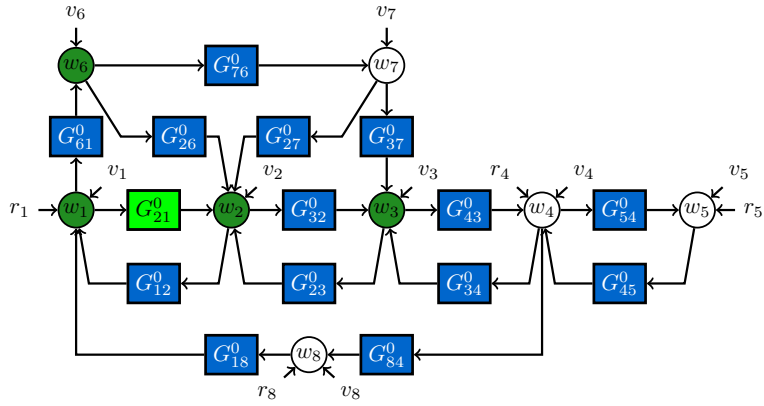

\medskip
When applying this result to the situation of the dynamic network in Figures \ref{figure:1} and \ref{fig.2sinput}, we can choose $\mathcal{D}_j=\{1,3,6\}$ as predictor inputs, while all reference signals can be used for projection, i.e. $\mathcal{T}_j= \{1,4,5,8\}$. The fact that $v_7$ is a confounding variable is no limitation now, and thus $w_7$ does not need to be included as a predictor input, and so does not need to be measured. The signals that need to be measured are indicated in green in Figure \ref{fig.2sinput}.

\section{Module identification with sensor noise}
It is often the case that the variables $w_k$ in (\ref{eq:netw_def}) are measured using sensors. The measurement error that results from imperfect sensor readings is called \emph{sensor noise} and denoted $s_k$. The measured version of the internal variable $w_k$ is modelled as
\begin{align*}
\tilde{w}_k(t) = w_k(t) + s_k(t),
\end{align*}
where $s_k$ is modelled as a stationary stochastic noise process with rational power spectral density.
In the open- and closed-loop system identification literature it is well known that sensor noise on the input can lead to biased estimates of the system dynamics if it is not properly handled \citep{Soderstrom12}. With input noise the identification problem turns into an \emph{errors-in-variables} problem. Because in a dynamic network setting an internal variable can often serve as both an "input" and an "output" to the identification problem, it is important to handle sensor noise properly to avoid unwanted bias in the estimated dynamics.

The main advantage of a dynamic network setting is that there may be opportunities to measure many variables, other than those strictly needed to identify the module of interest. These extra measurements can be used to handle the sensor noise.
The additionally selected internal and external variables can be used as \emph{instrumental variables (IV)}. Let $z$ denote a vector of instrumental variables:
\begin{align*}
z^T(t) = [\tilde{w}_{\ell_1}(t) \ \cdots \ \tilde{w}_{\ell_{n}}(t) \ r_{m_1}(t) \ \cdots \ r_{m_p}(t) ]
\end{align*}
and let $\mathcal{I}_j$ denote the set of indices $\{\ell_1,\cdots \ell_n\}$ of internal variables selected as instrumental variables, satisfying that $\mathcal{I}_j \cap \{\mathcal{D}_j,j\} = \emptyset$.

The main assumptions that we make are:
\begin{itemize}
\item all sensor noise terms $s_1, \ldots, s_L$ are mutually uncorrelated, i.e. the vector process $s$ has a diagonal spectral density, and
\item the sensor noises are uncorrelated to the process noises.
\end{itemize}

The mechanism to handle the sensor noise is to cross-correlate the predictor inputs with additionally measured internal variables, by considering
\[ R_{\tilde w_kz}(\tau) := \mathbb{E} \tilde w_k(t)z(t-\tau). \]
Due to the above two assumptions, the sensor noise is not present in $R_{\tilde w_kz}(\tau)$ for $\tau \ge 0$, leading to $R_{\tilde w_kz}(\tau)=R_{w_kz}(\tau)$. If there is a path from the instrumental variables to the predictor inputs $w_k$, $k \in \mathcal{D}_j$ then $R_{w_kz}(\tau)$ will be non-zero and will contain all the information of the dynamics of the network, while not being a function of the sensor noise.

Rather than considering the prediction error $\varepsilon(t,\theta)$ as in (\ref{eq:pe1}), we now construct the correlation between prediction error and instrumental variable:
\[ R_{\varepsilon z}(\tau) = H_j(\theta)^{-1}[R_{\bar w_j z}(\tau)-
   \sum_{k\in \mathcal{D}_j} G_{jk}(\theta) R_{w_k z}(\tau)], \]
and we replace the quadratic identification criterion:
\[ \hat\theta_N = \arg\min_{\theta\in\Theta} \sum_{\tau=0}^{n_z} R^2_{\varepsilon z}(\tau). \]

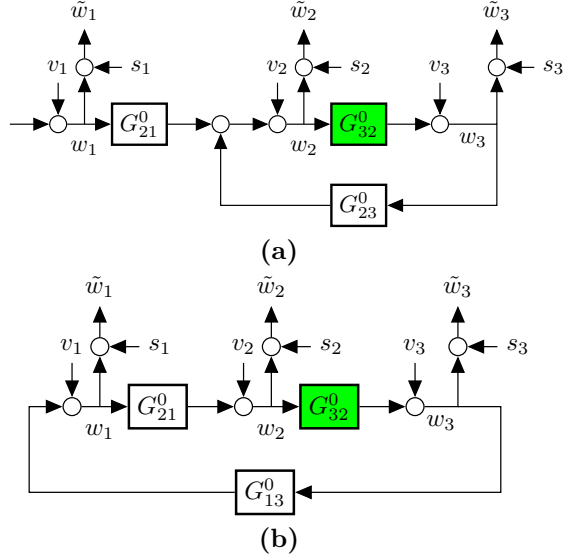
\begin{figure}[bht]
\begin{center}
\scalebox{0.90}{
\begin{tikzpicture}
[tf/.style={rectangle,draw,minimum height=0.4cm, line width=1pt, minimum width=0.5cm},
 sum/.style={circle,draw,minimum size=0.25cm, line width=0.5pt, inner sep=0cm},
 node distance = 0.6cm, line width = 0.5pt, >=triangle 45]
\node (x) {};
\node[sum] (v1sum) at ($(x)+0.7*(\dd,0)$){};
\node[tf] (g21) at ($(v1sum) + (\dd,0)$){$G_{21}^0$};
\node[sum] (loopSum) at ($(g21) + (\dd,0)$) {};
\node[sum] (v2sum) at ($(loopSum) + 0.7*(\dd,0)$) {};
\node[tf,fill=green] (g32) at ($(v2sum)+(\dd,0)$) {$G_{32}^0$};
\node[tf] (g23) at ($(g32)+(0,-\dd)$) {$G_{23}^0$};
\node[sum] (v3sum) at ($(g32) + (\dd,0)$) {};

\node (w1) at ($0.5*(v1sum)+0.5*(g21.west)$) [label=below:$w_1$] {};
\node[sum] (n1sum) at ($(w1) + 0.7*(0,\dd)$) {};
\node (n1) at ($(n1sum)+(0.7*\dd,0)$) {${s}_1$};
\node (m1) at ($(n1sum)+0.7*(0,\dd)$) {$\tilde{w}_1$};
\node (w2) at ($0.5*(v2sum)+0.5*(g32.west)$)[label=below:$w_2$]{};
\node[sum] (n2sum) at ($(w2)+0.7*(0,\dd)$){};
\node (n2) at ($(n2sum)+(0.7*\dd,0)$) {${s}_2$};
\node (m2) at ($(n2sum)+0.7*(0,\dd)$) {$\tilde{w}_2$};
\node (w3) at ($(v3sum)+0.7*(\dd,0)$){};
\node[sum] (n3sum) at ($(w3) + 0.7*(0,\dd)$) {};
\node (n3) at ($(n3sum) + (0.7*\dd,0)$) {${s}_3$};
\node (m3) at ($(n3sum) + 0.7*(0,\dd)$) {$\tilde{w}_3$};

\node (v1) at ($(v1sum) + (0,0.7*\dd)$) {$v_1$};
\node (v2) at ($(v2sum) + (0,0.7*\dd)$) {$v_2$};
\node (v3) at ($(v3sum) + (0,0.7*\dd)$) {$v_3$};

\draw[->] (x) -- (v1sum);
\draw[->] (v1sum) -- (g21);
\draw[->] (g21) -- (loopSum);
\draw[->] (loopSum) -- (v2sum);
\draw[->] (v2sum) -- (g32);
\draw[->] (g32) -- (v3sum);
\draw[->] (v3sum) -- node[below]{$w_3$} ($(w3)$) -- (n3sum);
\draw[->] ($(w1)$) -- (n1sum);
\draw[->] (n1) -- (n1sum);
\draw[->] (n1sum) -- (m1);
\draw[->] ($(w2)$) -- (n2sum);
\draw[->] (n2) -- (n2sum);
\draw[->] (n2sum) -- (m2);
\draw[->] (n3) -- (n3sum);
\draw[->] (n3sum) -- (m3);
\draw[->] ($(w3)$) |- (g23);
\draw[->] (g23) -| (loopSum);
\draw[->] (v1) -- (v1sum);
\draw[->] (v2) -- (v2sum);
\draw[->] (v3) -- (v3sum);
\end{tikzpicture}}

{\bfseries (a)}

\scalebox{0.95}{
\begin{tikzpicture}
[tf/.style={rectangle,draw,minimum height=0.4cm, line width=1pt, minimum width=0.5cm},
 sum/.style={circle,draw,minimum size=0.25cm, line width=0.5pt, inner sep=0cm},
 node distance = 0.6cm, line width = 0.5pt, >=triangle 45]
\node[sum] (v1sum) {};
\node[tf] (g21) at ($(v1sum)+(\dd,0)$) {$G_{21}^0$};
\node[sum] (v2sum) at ($(g21) + (\dd,0)$) {};
\node[tf,fill=green] (g32) at ($(v2sum)+(\dd,0)$) {$G_{32}^0$};
\node[sum] (v3sum) at ($(g32) + (\dd,0)$) {};
\node[tf] (g13) at ($0.5*(v1sum) +0.5*(v3sum) + (0.25*\dd,-\dd)$) {$G_{13}^0$};

\node (w1) at ($0.5*(v1sum)+0.5*(g21.west)$) [label=below:$w_1$] {};
\node[sum] (n1sum) at ($(w1) + 0.7*(0,\dd)$) {};
\node (n1) at ($(n1sum)+(0.7*\dd,0)$) {${s}_1$};
\node (m1) at ($(n1sum)+0.7*(0,\dd)$) {$\tilde{w}_1$};
\node (w2) at ($0.5*(v2sum)+0.5*(g32.west)$)[label=below:$w_2$]{};
\node[sum] (n2sum) at ($(w2)+0.7*(0,\dd)$){};
\node (n2) at ($(n2sum)+(0.7*\dd,0)$) {${s}_2$};
\node (m2) at ($(n2sum)+0.7*(0,\dd)$) {$\tilde{w}_2$};
\node (w3) at ($(v3sum)+0.5*(\dd,0)$){};
\node[sum] (n3sum) at ($(w3) + 0.7*(0,\dd)$) {};
\node (n3) at ($(n3sum) + (0.7*\dd,0)$) {${s}_3$};
\node (m3) at ($(n3sum) + 0.7*(0,\dd)$) {$\tilde{w}_3$};

\node (v1) at ($(v1sum) + (0,0.7*\dd)$) {$v_1$};
\node (v2) at ($(v2sum) + (0,0.7*\dd)$) {$v_2$};
\node (v3) at ($(v3sum) + (0,0.7*\dd)$) {$v_3$};

\draw[->] (g13) -| ($(v1sum)+(-0.5*\dd,0)$) -- (v1sum);
\draw[->] (v1sum) -- (g21);
\draw[->] (g21) -- (v2sum);
\draw[->] (v2sum) -- (g32);
\draw[->] (g32) -- (v3sum);
\draw[->] (v3sum) -- node[below]{$w_3$} ($(w3)$) -- (n3sum);
\draw[->] ($(w1)$) -- (n1sum);
\draw[->] (n1) -- (n1sum);
\draw[->] (n1sum) -- (m1);
\draw[->] ($(w2)$) -- (n2sum);
\draw[->] (n2) -- (n2sum);
\draw[->] (n2sum) -- (m2);
\draw[->] (n3) -- (n3sum);
\draw[->] (n3sum) -- (m3);
\draw[->] ($(w3)$) -- ++(0.5*\dd,0) |- (g13);
\draw[->] (v1) -- (v1sum);
\draw[->] (v2) -- (v2sum);
\draw[->] (v3) -- (v3sum);
\end{tikzpicture}}

{\bfseries (b)}
\end{center}
\caption{Closed loop data generating systems with sensor noise \citep{Dankers&etal_Autom:15}.\label{fig.closedLoop}}
\end{figure}

In \cite{Dankers&etal_Autom:15} it is shown that when using this quadratic criterion in the algorithm of the Direct Method presented in Section \ref{secdir}, this leads to a consistent estimate of $G^0_{ji}$, provided that the following additional conditions are satisfied:
\begin{itemize}
\item The "predictor inputs", $R_{w_kz}(\tau)$, $\tau \ge 0$, $k \in \mathcal{D}_j$ must be sufficiently informative, and
\item There must be a delay in the paths from $w_j$ to the instrumental variables $w_{\ell}$, $\ell \in \mathcal{I}_j$.
\end{itemize}

When considering the dynamic networks in Figure \ref{fig.closedLoop}, while the objective is to estimate $G^0_{32}$, the classical instrumental variable (IV) method would work for situation (a), where $\tilde w_1$ could be chosen as an instrumental variable in a typical linear regression scheme, that requires the IV-signal to be correlated to the input signal (in this case $w_2$), but uncorrelated to the output noise $v_3$ \citep{Gilson&VandenHof:05}. In the system depicted in Figure \ref{fig.closedLoop}(b) this latter condition is not satisfied. However the dynamic network method discussed in this section would also work in case (b), when choosing $\tilde w_1$ as IV-signal, provided that there is a delay in $G^0_{13}$.

The overall observation is that in a dynamic network, sensor noise is more easily dealt with than in a simple open-loop or closed-loop system, simply because of the presence of multiple signals that can be used to extract the relevant information.

\section{Network identifiability}
When we move from the (local) identification of a single module to the (global) identification of either the dynamics or the topology of the full network, other questions have to be addressed. It also has to be noted that local identification of a module through estimating a MISO model, can only be justified (from a minimum variance perspective) if the process noises on the different node signals are uncorrelated.

In handling the full network, we will expand the assumption on the process noise, to allow for correlation over the different nodes, thus allowing $\Phi_v(\omega)$ to be non-diagonal.

When identifying the full network dynamics in that situation, one of the important questions is whether the dynamic network is uniquely represented in the model set that is chosen. E.g. if we allow transfer functions to appear between all node signals available, can we then uniquely identify the network? It can be expected that the answer to this question will be related to the type of excitation that is present in the network (on which locations are external excitation signals present?) and structural restrictions that we can impose on the chosen model set (which node signals are connected with each other?).

We will consider a network model set, to be defined as
\[ \mathcal{M} := \{ (G(\theta), H(\theta), R(\theta)), \theta\in\Theta \} \]
where a network model is represented by the triplet $M(\theta) = (G(\theta), H(\theta), R(\theta))$.

The network transfer function is denoted by
\[ T^0 := [I-G^0]^{-1} \begin{bmatrix} H^0 & R^0\end{bmatrix}, \]
reflecting the mapping from external signals $e$ and $r$ to node signals $w$. $T^0$ is the (MIMO) transfer function that can generally be uniquely identified from measured data, provided that the external excitation signals are sufficiently informative. The network transfer function of a particular model, represented by $\theta$, will be indicated by $T(\theta)$.

The concept of network identifiability addresses the question whether two different models in $\mathcal{M}$ have the same network transfer function. The model set $\mathcal{M}$ will be called {\it network identifiable} \citep{Weerts&etal_SYSID:15}, if for any two models $M(\theta_1)$ and $M(\theta_2)$ in $\mathcal{M}$ it holds that
\[ \{T(\theta_1) = T(\theta_2)\} \Longrightarrow \{M(\theta_1) = M(\theta_2)\}. \]

Note that network identifiability is referring to a property of the model set in terms of properties of the models $M$. This is in contrast with the classical identifiability concept that is typically formulated in terms of uniqueness of parameter values \citep{ljung:99}. The model set $\mathcal{M}$ can be chosen in different ways, depending on the structural conditions that we would like to impose on the models. If all possible modules (connecting each and every node in the network) are parametrized, then identifiability of the model set will require strong conditions on the excitation signals being present.

Under rather generic conditions on the model set (see \cite{weerts&etal_ArXiv:16}), network identifiability can be shown if the following conditions are satisfied:
\begin{enumerate}
	\item Each row $i$ of $\begin{bmatrix}G(\theta) & H(\theta) & R(\theta) \end{bmatrix}$ has at most $K+p$  parameterized entries, where $K$ is the number of external excitation signals and $p$ is the number of white noise processes driving the process noise $v$, and
	\item For each row $i$, $\breve T$ has full row rank, where $\breve T$ is the submatrix of $T$ composed of those rows $j$ that correspond to elements $G_{ij}(\theta)$ that are parameterized, and of those columns $k$ that correspond to elements $U_{ik}(\theta)$ in $U(\theta):=\begin{bmatrix} H(\theta) & R(\theta) \end{bmatrix}$ that are not parameterized.
\end{enumerate}
The second condition is automatically satisfied whenever $\begin{bmatrix} H(\theta) & R(\theta) \end{bmatrix}$ is of diagonal structure and has full row rank for all $\theta$, being a situation which is guaranteed if each and every node signal has either an excitation signal $r$ or an independent process disturbance $v$ directly connected to it.

\begin{figure}[h]
\centering
\includegraphics[width=0.95\linewidth]{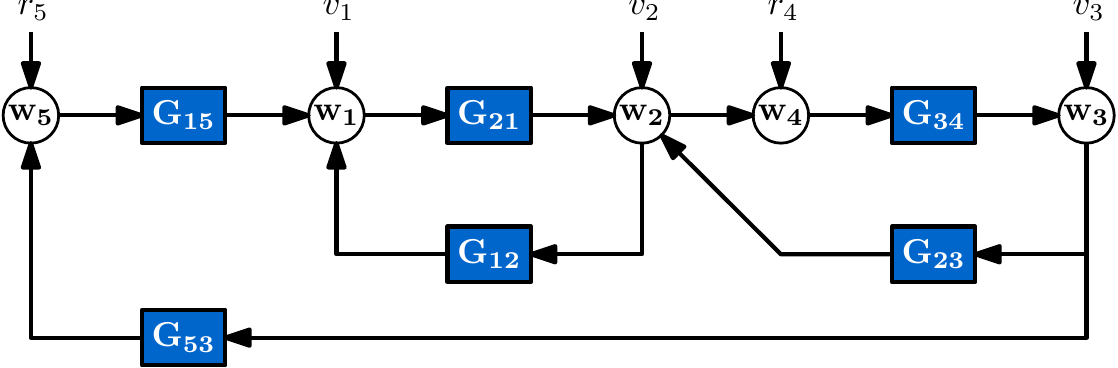}
\caption{Five node dynamic network \citep{weerts&etal_ArXiv:16}.}\label{figure:5}
\end{figure}

As an example consider the dynamic network in Figure \ref{figure:5}, with three process noises and two external excitations. We can consider three different situations now:
\begin{enumerate}
\item If the three noise terms are known to be uncorrelated, and the presence of $r_4$ and $r_5$ is known, then
\[ \begin{bmatrix} H(\theta) & R(\theta) \end{bmatrix} =
   \begin{bmatrix} H_1(\theta) & 0 & 0 & 0 & 0 \\ 0 & H_2(\theta) & 0 & 0 & 0 \\
   0 & 0 & H_3(\theta) & 0 & 0 \\ 0 & 0 & 0 & 1 & 0 \\ 0 & 0 & 0 & 0 & 1 \end{bmatrix} \]
being diagonal and of full row rank, implying that for any parametrization of $G(\theta)$, the model set is identifiable. This is caused by the independent excitation of each node of the network.
\item If $v_1$ and $v_2$ are known to be correlated, they have to be modelled in a multivariable way, leading to
\[ \begin{bmatrix} H(\theta) & R(\theta) \end{bmatrix} =
   \begin{bmatrix} H_{11}(\theta) & H_{12}(\theta) & 0 & 0 & 0 \\ H_{21}(\theta) & H_{22}(\theta) & 0 & 0 & 0 \\
   0 & 0 & H_3(\theta) & 0 & 0 \\ 0 & 0 & 0 & 1 & 0 \\ 0 & 0 & 0 & 0 & 1 \end{bmatrix} \]
Since we do not have a guarantee anymore that this matrix is full row rank, we loose network identifiability if the network structure in $G(\theta)$ is fully parametrized. A fully parametrized $G(\theta)$ will be a $5 \times 5$ matrix with $4$ parametrized transfers on each row. This implies that the first two rows of the matrix $\begin{bmatrix}G(\theta) & H(\theta) & R(\theta) \end{bmatrix}$ will have $6$ parametrized entries, while the number of external signals $K+p = 5$. In this situation we loose identifiability.
\item If we consider the same situation, but now using information on the topology of the network, i.e. we parametrize
    \[ G(\theta) = \begin{bmatrix}
0 & \!G_{12}(\theta)\! & 0 & 0 &\! G_{15}(\theta)\\
G_{21}(\theta)\!&0&\!G_{23}(\theta)\!&0&0 \\
0&0&0&\!G_{34}(\theta)\!&0\\
0&1&0&0&0\\
0&0&\!G_{53}(\theta)\!&0&0
\end{bmatrix} \]
then the maximum number of parametrized entries in $\begin{bmatrix}G(\theta) & H(\theta) & R(\theta) \end{bmatrix}$ is $4$ and the first condition for network identifiability is satisfied. It can be verified that in this case also the second condition is satisfied, and that network identifiability is guaranteed.
\end{enumerate}

\section{The interacting two-control loops}

The situation of the interacting two control loops as sketched in Figure \ref{figure:3}, has been treated in detail in \cite{Gudi&Rawlings:06}, with respect to the identification of the interacting dynamics $G_{21}$ for the situation that $G_{12}=0$. When considering this problem in the scope of the network identification approaches in the present paper, the following observations can be made:
\begin{itemize}
\item A direct way to identify $G_{21}$ would be to model the two-input one-output system, having $u_1$ and $u_2$ as inputs and $y_2$ as output, and applying a direct identification method. Provided that $u_1$ and $u_2$ are sufficiently exciting, this would lead to a consistent estimate of $G_{21}$ and $G_2$, under the condition that an appropriate noise model is identified too.
\item Alternatively the necessity to include a noise model could be relaxed by using a two-stage / projection approach where the input signals $u_1$ and $u_2$ are projected onto the external excitation signals $r_1$ and $r_2$, as well as onto possible additional dither signals added to the plant inputs.
\end{itemize}

Given the results in this paper, also the situation of having both interacting dynamics $G_{21}$ and $G_{12}$ being present can be treated, by identifying two two-input one-output models, having $u_1$ and $u_2$ as inputs, and either $y_1$ or $y_2$ as outputs. From a variance point of view it would be attractive to combine those two identification tasks in the identification of one multivariable two-input two-output model.

\section{Discussion and challenges}

We have presented a schematic picture on identification methods and tools that are suitable for identification of modules in dynamic networks, that are based on classical closed-loop identification schemes.
Identification in/of dynamic networks is a challenging area for which many of the problems still need to be sorted out. Questions like: optimal locations of sensors and actuators in order to achieve a particular model accuracy, experiment design, and identifying the topology of networks on the basis of data, are just a few topics that can be mentioned in this sense. Additionally, while allowing to let the networks grow in dimensions, the scalability of algorithms will become an important topic.

\section*{Acknowledgments}
This work has received funding from the European Research Council (ERC), Advanced Research Grant SYSDYNET, under the European Union's Horizon 2020 research and innovation programme (grant agreement No 694504). The authors acknowledge the discussions with and contributions of Xavier Bombois, Peter Heuberger and Jobert Ludlage.

\section*{References}
\bibliographystyle{plainnat}		
\bibliography{FocapoLibrary}

\begin{thebibliography}{19}
\providecommand{\natexlab}[1]{#1}
\providecommand{\url}[1]{\texttt{#1}}
\expandafter\ifx\csname urlstyle\endcsname\relax
  \providecommand{\doi}[1]{doi: #1}\else
  \providecommand{\doi}{doi: \begingroup \urlstyle{rm}\Url}\fi

\bibitem[Christofides et~al.(2013)Christofides, Scattolini,
  Mu$\tilde{\mbox{n}}$oz de~la Pe$\tilde{\mbox{n}}$a, and
  Liu]{Christofides&etal:13}
P.~G. Christofides, R.~Scattolini, D.~Mu$\tilde{\mbox{n}}$oz de~la
  Pe$\tilde{\mbox{n}}$a, and J.~Liu.
\newblock Distributed model predictive control: A tutorial review and future
  research directions.
\newblock \emph{Computers and Chemical Engineering}, 51:\penalty0 21--41, 2013.

\bibitem[Dankers et~al.(2015)Dankers, Van~den Hof, Bombois, and
  Heuberger]{Dankers&etal_Autom:15}
A.~G. Dankers, P.~M.~J. Van~den Hof, X.~Bombois, and P.~S.~C. Heuberger.
\newblock Errors-in-variables identification in dynamic networks - consistency
  results for an instrumental variable approach.
\newblock \emph{Automatica}, 62:\penalty0 39--50, 2015.

\bibitem[Dankers et~al.(2016)Dankers, Van~den Hof, Heuberger, and
  Bombois]{Dankers&etal_TAC:16}
A.~G. Dankers, P.~M.~J. Van~den Hof, P.~S.~C. Heuberger, and X.~Bombois.
\newblock Identification of dynamic models in complex networks with predictior
  error methods - predictor input selection.
\newblock \emph{IEEE Trans. Automatic Control}, 61\penalty0 (4):\penalty0
  937--952, 2016.

\bibitem[Darby and Nikolaou(2014)]{Darby&Nikolaou:14}
M.~L. Darby and M.~Nikolaou.
\newblock Identification test design for multivariable model-based control: an
  industrial perspective.
\newblock \emph{Control Engineering Practice}, 22:\penalty0 165--180, 2014.

\bibitem[Forssell and Ljung(1999)]{Forssell&Ljung:99}
U.~Forssell and L.~Ljung.
\newblock Closed-loop identification revisited.
\newblock \emph{Automatica}, 35\penalty0 (7):\penalty0 1215--1241, 1999.

\bibitem[Gevers(2005)]{Gevers:05}
M.~Gevers.
\newblock Identification for control: from the early achievements to the
  revival of experiment design.
\newblock \emph{European J. Control}, 11:\penalty0 1--18, 2005.

\bibitem[Gevers and Bazanella(2015)]{Gevers&Bazanella:15}
M.~Gevers and A.~S. Bazanella.
\newblock Identification in dynamic networks: identifiability and experiment
  design.
\newblock In \emph{Proc. 2015 IEEE 54th Conf. Decision and Control, Osaka,
  Japan}, pages 4006--4011, 2015.

\bibitem[Gilson and Van~den Hof(2005)]{Gilson&VandenHof:05}
M.~Gilson and P.~M.~J. Van~den Hof.
\newblock Instrumental variable methods for closed-loop system identification.
\newblock \emph{Automatica}, 41\penalty0 (2):\penalty0 241--249, 2005.

\bibitem[Gudi and Rawlings(2006)]{Gudi&Rawlings:06}
R.~D. Gudi and J.~B. Rawlings.
\newblock Identification for decentralized model predictive control.
\newblock \emph{AIChE Journal}, 52\penalty0 (6):\penalty0 2198--2210, 2006.

\bibitem[Ljung(1999)]{ljung:99}
L.~Ljung.
\newblock \emph{System Identification: Theory for the User}.
\newblock Prentice-Hall, Englewood Cliffs, NJ, 1999.

\bibitem[\"{O}zkan et~al.(2016)\"{O}zkan, Bombois, Ludlage, Rojas, Hjalmarsson,
  Mod\'{e}n, Lundh, Backx, and Van~den Hof]{Autoprofit:16}
L.~\"{O}zkan, X.~Bombois, J.~H.~A. Ludlage, C.~Rojas, H.~Hjalmarsson, P.~E.
  Mod\'{e}n, M.~Lundh, T.~C. P.~M. Backx, and P.~M.~J. Van~den Hof.
\newblock Advanced autonomous model-based operation of industrial process
  systems (autoprofit): Technological developments and future perspectives.
\newblock \emph{Annual Reviews in Control}, 42:\penalty0 126--142, 2016.

\bibitem[Rawlings and Stewart(2008)]{Rawlings&Stewart:08}
J.~B. Rawlings and B.~T. Stewart.
\newblock Coordinating multiple optimization-based controllers: new
  opportunities and challenges.
\newblock \emph{Journal of Process Control}, 18:\penalty0 839--845, 2008.

\bibitem[S\"oderstr\"om(2012)]{Soderstrom12}
T.~S\"oderstr\"om.
\newblock System identification for the errors-in-variables problem.
\newblock \emph{Transactions of the Institute of Measurement and Control},
  34:\penalty0 780--792, 2012.

\bibitem[Van~den Hof and Schrama(1993)]{VandenHof&Schrama:93}
P.~M.~J. Van~den Hof and R.~J.~P. Schrama.
\newblock An indirect method for transfer function estimation from closed loop
  data.
\newblock \emph{Automatica}, 29\penalty0 (6):\penalty0 1523--1527, 1993.

\bibitem[Van~den Hof and Schrama(1995)]{VandenHof&Schrama:95}
P.~M.~J. Van~den Hof and R.~J.~P. Schrama.
\newblock Identification and control-closed loop issues.
\newblock \emph{Automatica}, 31\penalty0 (12):\penalty0 1751--1770, 1995.

\bibitem[Van~den Hof et~al.(2013)Van~den Hof, Dankers, Heuberger, and
  Bombois]{VandenHof&etal:13}
P.~M.~J. Van~den Hof, A.~G. Dankers, P.~S.~C. Heuberger, and X.~Bombois.
\newblock Identification of dynamic models in complex networks with prediction
  error methods - basic methods for consistent module estimates.
\newblock \emph{Automatica}, 49\penalty0 (10):\penalty0 2994--3006, 2013.

\bibitem[Weerts et~al.(2015)Weerts, Dankers, and Van~den
  Hof]{Weerts&etal_SYSID:15}
H.~H.~M. Weerts, A.~G. Dankers, and P.~M.~J. Van~den Hof.
\newblock Identifiability in dynamic network identification.
\newblock \emph{IFAC-PapersOnLine}, 48\penalty0 (28):\penalty0 1409--1414,
  2015.
\newblock Proc. 17th IFAC Symp. System Identification, Beijing, China.

\bibitem[Weerts et~al.(2017)Weerts, Van~den Hof, and
  Dankers]{weerts&etal_ArXiv:16}
H.~H.~M. Weerts, P.~M.~J. Van~den Hof, and A.~G. Dankers.
\newblock Identifiability of dynamic netwoks with noisy and noise-free nodes.
\newblock arXiv:1609.00864[cs.SY], provisionally accepted for publication in
  Automatica, 2017.

\bibitem[Zhu(2006)]{Zhu:06}
Y.C. Zhu.
\newblock System identification for process control: Recent experience and
  outlook.
\newblock In \emph{IFAC Proceedings Volumes}, volume 9(1), pages 20--32, 2006.
\newblock Proc. 14th IFAC Symp. System Identification.

\end{thebibliography}

\end{document}